\documentclass[12pt]{iopart}
\usepackage{iopams}
\usepackage{graphicx,color}
\usepackage{hyperref}
\usepackage{braket}
\usepackage[toctitles]{titlesec}
\begin{document}

\title [Spherical model on a   spider-web graph] { Non-perturbative corrections to mean-field critical behavior: Spherical model on a   spider-web graph}

\author{Ajit C Balram$^1$, Deepak Dhar$^2$}

\address{$^1$Department of Physics,
Indian Institute of Science Education and Research, \\
Sai Trinity Building,
Garware Circle, Sutarwadi, Pashan,
Pune  411 021, India}

\address{$^2$Department of Theoretical Physics, Tata Institute of
Fundamental Research,\\
1 Homi Bhabha Road, Mumbai  400 005, India}
%shorttitle{Spherical model on a spider-web graph}
\ead{\mailto{ddhar@theory.tifr.res.in}}

\begin{abstract}
We consider the spherical model on   a    spider-web graph.  This graph  is effectively infinite-dimensional, similar to the Bethe lattice,  but has loops. We show that these lead to  non-trivial corrections to the simple mean-field behavior.  We first determine all normal modes of the coupled springs problem on this graph, using its large symmetry  group. In the  thermodynamic limit,  the spectrum is a set of $\delta$-functions, and  all the modes are localized. The fractional number of modes with frequency less than $\omega$ varies as $\exp ( -C/\omega)$ for $\omega$ tending to zero, where $C$ is a constant. For an unbiased random walk on the vertices of this graph, this implies that the probability of return to the origin at time $t$ varies as $\exp( - C' t^{1/3})$, for large $t$, where $C'$ is  a constant. For the spherical model, we show that while the critical exponents take the  values expected from the  mean-field 
theory, the free-energy per site at temperature $T$, near and above the critical temperature $T_c$,  also has an essential singularity of the type $\exp[ -K {(T - T_c)}^{-1/2}]$.

\end{abstract}

\submitto{\JPA}
\maketitle

\section{ Introduction}

It is well-known that in systems having dimension $d$ greater than the upper critical dimension $d_U$, fluctuations in thermal equilibrium near  critical phase transitions are well described by the mean-field theory \cite{textbook}.  For simple Ising-like spin models, the value of upper critical dimension is known to be $4$.  In this paper, we give an example of a lattice, whose effective dimension is infinite, and the exact solution of the spherical model on the lattice shows that while the critical exponents  are correctly given by the mean-field theory, (e.g. the specific heat has a discontinuity),  the free-energy near the transition also shows an essential singularity, with the singular part varying as  $\exp[ -K {(T - T_c)}^{-1/2}]$, for temperatures $T$ near and greater than the critical temperature $T_c$. While  it is known that higher derivatives of the free energy do show logarithmic singularities in the spherical model for even dimensions $d > d_U$ \cite{joyce},  this seems to be the first case where an essential singularity is encountered in the sub-leading singular part of the  free energy   in an infinite dimensional model.

The lattice we study is the spider-web lattice, also known as spider-web graph or the spider-web network in  literature. These graphs  were introduced by Ikeno in 1959, for efficient design of crossbar exchanges for telephone switching networks \cite{Ikeno59}. The name spider-webs to describe them  seems to have been first used  by Feiner and Kappel in 1970 \cite{Feiner_Kappel}. With the advent of digital switching and mobile communication networks, crossbar exchanges have become obsolete, and their optimal design  is no longer of  technological interest. But the spider-web graphs are very interesting structures for statistical physics.

A $2^N\times M$ - node spider network graph  consists of  $M$-levels each of which has $2^N$ vertices.  Each  vertex in the $m$-th level   is connected to two vertices each  in the $(m \pm 1)$th levels. 
These  are finite graphs where all vertices are equivalent.  In the thermodynamic limit when the number of vertices in the graph tends to infinity,  the number of sites within a distance $R$ of a given site increases exponentially with $R$, and the effective dimension of these graphs is infinite. These are thus similar to the Bethe lattice, which has been studied a lot in statistical physics. However, for the Bethe lattice, realized as the limit of tree graphs, separating the bulk properties from the surface properties requires care as most of its sites are near the surface. One can realize a Bethe lattice without surface effects as the limit of a random graph with uniform coordination number, but this involves  averaging over disorder, which may be quenched \cite{quench}, or annealed \cite{annealed}.  Spider-web graphs avoid the difficulties associated with such  averagings.  They also have additional symmetries, in addition to translational invariance, which  makes the study of statistical physics models on these graphs analytically more tractable.

Given its origins, it is not surprising  that most of the earlier studies  dealt with the percolation properties of these networks \cite{Ikeno59}. Takagi {\cite{Takagi68}} showed that in a certain class of graphs, the spider-web network has the highest linking probability. But, Chung and Hwang \cite{Chung_Hwang80} showed  that,  outside this class, there are other graphs that are better. For subsequent work on the percolation properties on these networks, see \cite{Hwang82, Pippenger91, Pippenger92, Pippenger06}. Other models on these graphs have not been studied much so far.  There is large amount work dealing with spectra of different graphs. Some basic references may be found in \cite{chung}.

In this paper, we  study the spherical model  on the spider-web graph. We start by  determining the normal modes of vibration of masses  coupled by nearest -neighbor  harmonic springs on this network. We obtain the entire spectrum for finite  values of $M$ and $N$, and also describe the qualitative features of the spectrum in the thermodynamic limit of $M\rightarrow \infty$, with $N$ finite.   In the double limit of both $M, N \rightarrow \infty$, almost all the eigenmodes of the springs problem are localized on this network. We study the return to origin for the related problem of a random walker on this network.  We show that the probability that a random walker returns to origin after $t$ steps  varies as $\exp(- t^{1/3})$. This should be compared with a power-law decay $t^{-d/2}$ on $d$-dimensional hypercubic lattices, and an exponential decay   on the Bethe lattice.  We use  the  knowledge of normal modes of the springs model on this lattice to determine the 
critical behavior of the spherical model on this graph, and show that the free energy shows an essential singularity at the transition point.

The plan of this paper is as follows.  In section II, we define spider-web graphs, and then define the springs Hamiltonian, the random walk problem  and the spherical model on these graphs. In section III, we discuss the decomposition of the Hilbert space of the displacement field in the springs problem into subspaces, called sectors, using the symmetries of the spider-web graphs. In Section IV, we determine the frequencies of all the normal modes of the springs Hamiltonian, and  also determine the  eigenmodes. In Section V, we use this spectrum  to determine the exact generating function for returns to origin of a walker doing a random walk on the vertices of the spider web graph. In section VI, we show that the  singular nature of density of states of the springs model  for small frequencies implies that free energy of the spherical model shows a discontinuity in specific heat, as expected from the standard mean-field theory, but also an  essential singularity.

\section{Definition of the model}

A spider-web network is a graph consisting  of $M q^N$ vertices, divided into $M$ levels, each level with  $q^N$ vertices, with $M, N$ and $q$ being positive integers.  We label the levels by integers $1$ to $M$, and  vertices within a level are labeled by $N$-digit integers $r$, base $q$, with $0 \leq r \leq (q^N -1)$.  In the following we discuss only the case $q=2$. Generalization to other values of $q$ is straight-forward.

On a binary string of length $N$, we define an operation $R_0$ of adding a zero from the right, making space for it by shifting all bits by one space to the left. The left-most (first)  bit of the original string  is lost.  So, for example, for $N=6$, action of $R_0$ on the binary string $111001$ yields a string $110010$, and we write $R_0(111001)= 110010$. The operator $R_1$ adds a $1$ to the right.  We similarly define operators $L_0$ and $L_1$, which correspond to adding  $0$ and $1$ respectively at the left end, and shifting the bits  one space to the right. In this case, the rightmost bit is lost.
Clearly, $R_0(r) = 2 r ( mod ~2^N)$, and $R_1(r) = 2 r + 1 ( mod ~2^N)$. Also,  $L_0(r)= \lfloor r/2 \rfloor$ and $L_1(r)= 2^{N-1} +\lfloor r/2 \rfloor$.

\begin{figure}
\begin{center}
\includegraphics[width=4.0in]{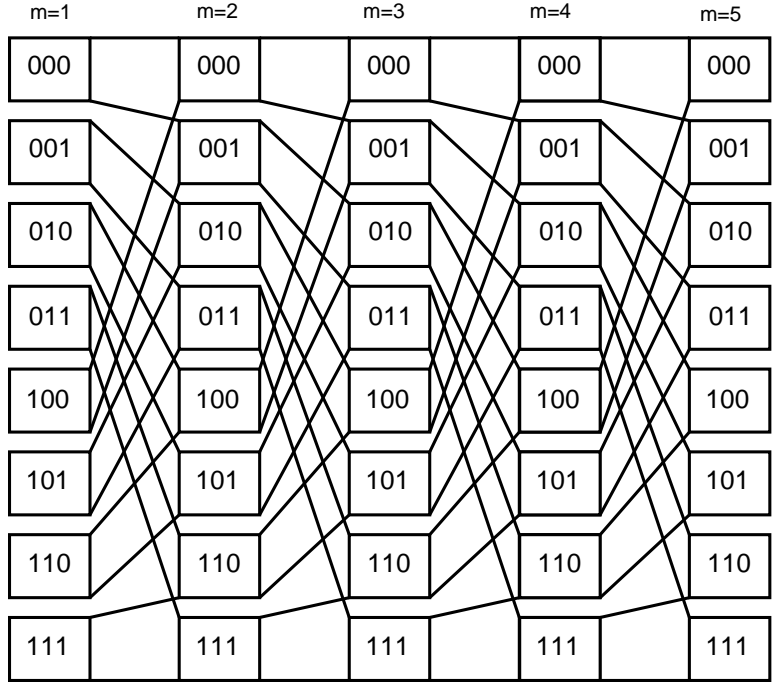}
\caption{$N=3, M=4$ spider network with periodic boundary conditions, the vertices in the column $m=5$ are identified with those of $m=1$. A vertex with $3$-bit label $\ell$ in the level $m$ is connected to the vertices with labels $R_0(\ell)$, and $R_1(\ell)$ in the level $(m+1)$. }
\label{fig:1}
\end{center}
\end{figure}

Now consider a vertex labeled by a binary integer $r$ in the level $m$. We denote this vertex by $(m,r)$.
The spider-web graph is defined by connecting this vertex to the two vertices $(m+1,R_0(r))$ and $(m+1,R_1(r))$ in the level $(m+1)$.  We assume periodic boundary conditions, and the vertices in the level $m=M$ are connected to level $m=1$. It is convenient to assume that  the index $m$ is defined only modulo $M$, so that $m=1$ is the same as $m=M+1$. We will take these connecting edges to be undirected. Then, each vertex $(m,r)$ is also connected to the
vertices $(m-1, L_0(r))$ and $(m-1, L_1(r))$.
We will assume that $M \geq N+1$, as otherwise the graph is not fully connected.  Fig. \ref{fig:1} shows a graph with $M=4, N=3$.

Given this structure of connections, it is easy to see that there is unique directed path of length $N$ from any vertex $(m,r)$ to any of the $2^N$ vertices in the level $(m+N)$. If the binary label of the target vertex is $i_1 i_2 i_3 ...i_N$, one just starts at the vertex $(m, r)$, and takes steps to the next level using edges corresponding to $R_{i_1}, R_{i_2}, R_{i_3}...$
respectively.

In the context of crossbar exchanges, we think of each node of the graph as $2 \times 2$ crossbar, with two input lines, and two output lines. It is natural to work with open boundary conditions, with no direct connections between level $M$ and level $1$.  Then there are $2^{N+1}$ input lines coming into level $1$, and there is an equal number of lines coming out of level $M$. The total numer of crosspoints used is $4$ per vertex of graph.  Then, using  only $M = N$ levels,  we can connect any of $2^{N+1}$  input lines into the  level $m=1$, we can connect to  any of the $2^{N+1}$ lines of the crossbars in the level $m=N$.  Also, there is a simple algorithm to determine the path from the  input node to the output node.  This is much less than  the number $2^{N+1} \times 2^{N+1}$, which would have been required, if a single stage crossbar was used.  If $M > N$, we get multiple possible paths from the input to output. This was the original reason
for  interest in these networks.

For studies in statistical physics, one is usually interested in the thermodynamical limit, where the effect of 
boundaries can be ignored. We note that here, local properties of the network do not depend on $M$ or $N$, for 
distances up to $N$.  For example, the number of sites that are within a distance $r$  of any site is the same for all sites, and is independent of $M$ and $N$,  so long as  $ N > r$. Thus, for systems with  interactions of finite range, say for a nearest-neighbor Ising model defined on this graph, the thermodynamical limit will be well-defined.  As another example, if there is a random walker that moves on the vertices of this graph, taking a step at random along one of the four edges at the current position, the probability of return
to starting point after $T$ steps is well defined, and independent of $M$ and $N$ , so long as $ T <  2 N$.

We consider a unit mass attached to each vertex of the spider-web network, coupled to its four neighbors by harmonic springs of unit spring constant. We denote a scalar variable $x({\bf v})$ for each vertex $ {\bf v} \equiv (m,r)$ of the graph. Let the corresponding momentum variable be denoted by $p({\bf v})$. Then,
the  classical -mechanical springs Hamiltonian is given by

\begin{equation}
H =  \frac{1}{2}\sum_{{\bf v}} p({\bf v})^2+
\frac{1}{2}\sum_{{\bf v},{\bf v'}} x({\bf v})\mathbb {K}({{\bf v}, {\bf v'}}) x({\bf v'})
\label{eq1}
\end{equation}

where  $\mathbb{K}$ is  the matrix of coupling constants defined by
\begin{eqnarray}
\mathbb{K}({\bf v},{\bf v'}) &=& 4, {\rm if} ~{\bf v}={\bf v'}, \nonumber \\
&=& -1, {\rm ~if} ~{\bf v}, {\bf v'} {\rm  ~are ~~nearest~~ neighbors},\nonumber \\
 &=&0, {\rm ~otherwise}
\end{eqnarray}

Clearly, the matrix ${\mathbb K}$ can be expressed in terms of the adjacency matrix of the graph.
As is well-known, the problem of determining the eigenmodes of the set of coupled oscillators reduces
to diagonalizing the interaction matrix ${\mathbb K}$, equivalently finding the spectrum of the discrete Laplacian of the graph. Let $\psi_{\alpha} ({\bf v})$ be a normal mode of this system with frequency $\omega_{\alpha}$. Then, $\psi_{\alpha}$
is a solution of the eigenvalue equation

\begin{equation}
\mathbb{K} \psi_{\alpha}({\bf v})=  \omega_{\alpha}^2 \psi_{\alpha}({\bf v}).
\label{eq3}
\end{equation}

We will denote by $F(\omega^2)$, the fractional number of modes of  an eigenvalue less than or equal to $\omega^2$. Hence, $0\leq F(\omega^2) \leq 1$. This is called the  cumulative spectral density.  The spectral density $D(\omega^2)$ is defined as $dF(\omega^2)/d\omega^2$. In the limit $M \rightarrow \infty$, with $N$ finite, we will show that $D(\omega^2)$ has both a discrete part, and a continuous part. The fraction  of modes that give the continuous part is $2^{-N}$. Therefore, in the limit  $N \rightarrow \infty$, we  get only  a discrete spectrum.

  It is well-known that the diagonalization of the springs Hamiltonian is also related to problem of random walks.
Consider a continuous-time random walk on the vertices of this graph, where  the walker can jump to any one of the  neighboring  vertices  with rate $1$.  Thus, in a short time $dt$, it can jump to any of the four neighbors with probability $dt$, and stay at the same vertex with probability $1- 4 dt$. Suppose the walker starts at vertex ${\bf v}_0$. Then, we denote the probability that the walker is at vertex ${\bf v}$ at time $t$,  by ${\rm Prob}({\bf v},t)$. We define a vector $|P(t)\rangle$, whose ${\bf v}$-th component is ${\rm Prob}({\bf v},t)$. Then the evolution of $|P(t)\rangle$  satisfies the Markov  equation

\begin{equation}
\frac{d}{dt}|P(t)\rangle =  -  {\mathbb K} |P(t)\rangle
\end{equation}

Then, the probability of being at vertex ${\bf v'}$ at time $t$ is given by

\begin{equation}
{\rm Prob}({\bf v'},t|{\bf v}_0,0) =  \langle {\bf v'} | e^{- {\mathbb K}  t}|{\bf v}_0\rangle.
\label{eq:5}
\end{equation}

The probability of being at the starting point at time $t$ is given by ${\rm Prob}({\bf v}_0,t|{\bf v}_0,0)$.
From the translational symmetry of the graph,  this does not depend on ${\bf v}_0$. We define the
Laplace -transform of  return probabilities $R(s)$ by
\begin{equation}
R(s) = \int_0^{\infty} dt  ~{\rm Prob}({\bf v}_0,t|{\bf v}_0,0) e^{-s t}
\end{equation}

Then it easy to see that
\begin{equation}
R(s) = \frac{1}{M 2^N} Tr [ s  +  {\mathbb K}]^{-1} = \int_0^{\infty} d\omega^2 D(\omega^2) [ s + \omega^2]^{-1}
\label{eq:7}
\end{equation}

We use this determination of the normal modes to solve the spherical model on this graph. In the spherical model, we assign a scalar variable $s_{\bf v}$ to each vertex ${\bf v}$ of the graph. These variables are subjected to a single constraint, known as the spherical constraint

\begin{equation}
\sum_{\bf v} s_{\bf v}^2 =  N_s
\label{sph_constr}
\end{equation}
where $N_s$ is the total number of sites in the graph. The spherical model can be obtained as the large $n$ limit of an $n$-vector model on the same graph \cite{joyce}, where we assign an $n$-vector spin  $\vec{S}_{\bf v}$ to each 
site ${\bf v}$, with a constraint $|\vec{S}_{\bf v}|^2 = 1$, at each vertex ${\bf v}$. The hamiltonian of the spherical model is given by 

\begin{equation}
{\cal{H}}_{sph} = \frac{1}{2}  \sum_{{\bf v, v'}} s_{\bf{v}} ~ {\mathbb K}( {\bf v},{\bf v'}) ~ s_{\bf{v'}}
\end{equation}

\section{Symmetries and sector decomposition}

As noted before, the problem of determination of the normal modes of the springs hamiltonian $H$ reduces to that of  diagonalizing  the matrix ${\mathbb K}$. This is an $N_s \times N_s$ matrix, where $N_s$ is the number of vertices in the graph, $ N_s = M 2^N$.  The matrix ${\mathbb K}$ may be viewed as the discrete laplacian operator, acting on the vector space of square-integrable real   functions defined on the vertices of the graph. Using the symmetries of the problem, we can block-diagonalize the matrix ${\mathbb K}$. We will call
the subspaces  corresponding to different blocks as sectors.  

The spider-web network has a high degree of symmetry.  Firstly, all vertices are equivalent.  We define
a translation operator ${\mathbb T}$ which takes a vertex $(m,r)$ to $(m+1,r)$. Clearly, ${\mathbb T}$ is a symmetry
of the graph, and we have
\begin{equation}
{\mathbb T}^M = {\mathbb I}
\end{equation}
where ${\mathbb I}$ is the identity operator.

There are additional symmetries. Consider any integer $i$ lying  between $1$ and $M$.   Define $f_k(r)$ as the integer obtained from $r$ by flipping the $k$-th bit of its binary representation, reading from left to right.   If the bit is $0$, it is replaced by $1$, and vice versa. We define the   operators ${\mathbb F}_k$,  for $k = 1$ to $N$. The operator ${\mathbb F}_k$ takes the site $(m +k -1,r)$ to another site $(m +k -1,f_k(r))$, for all $r, m$.  Sites in levels $j$ where $m-j+1 (mod~M) $ is outside the range $1$ to $N$ are left unchanged.
Clearly, we have ${\mathbb F}_i^2 = {\mathbb I}$. Also, we see that this mapping preserves the adjacency strucure of the network: if vertices ${\bf v}$ and ${\bf v'}$ are nearest neighbors, so are ${\mathbb F}_i ({\bf v})$ and ${\mathbb F}_i ({\bf v'})$.
Hence we have
\begin{equation}
[{\mathbb K},{\mathbb F}_i]  =0.
\end{equation}
Acting on the space of wavefunctions, the action of the operators ${\mathbb F}_i$ is given by
\begin{equation}
{\mathbb F}_i \psi({\bf v}) = \psi( {\mathbb F}_i {\bf v})
\end{equation}

Now, clearly, the operators ${\mathbb T}$ and ${\mathbb F}_i, i=1$ to $M$ commute with ${\mathbb K}$:
\begin{equation}
[{\mathbb K},{\mathbb F}_i] = [{\mathbb K}, {\mathbb T}] =0.
\end{equation}

Also, different ${\mathbb F}_i$ commute with each other:
\begin{equation}
[ {\mathbb F}_i, {\mathbb F}_j] =0. {\rm ~ for ~~all ~}i,j.
\end{equation}

We define  shift-operator ${\mathbb R}_0$,  which acts on the set of vertices of the graph, and for all sites $(m,r)$, ${\mathbb R}_0 (m,r) = (m+1, R_0 (r))$. Operators ${\mathbb R}_1, {\mathbb L}_0, {\mathbb L}_1$ are defined similarly.
Then clearly, we have 
\begin{equation}
{\mathbb K} = {\mathbb L}_0 + { \mathbb L}_1 + {\mathbb R}_0 + {\mathbb R}_1 - 4
\end{equation}
We can think of ${\mathbb L}_0, {\mathbb L}_1, {\mathbb R}_0$ and $ {\mathbb R}_1$ as the generators a semigroup.  These can be written as $N_s \times N_s$ matrices.

If we first apply ${\mathbb R}_a$, and then ${\mathbb L}_b$ on any vertex ${\bf v}$, where $a$ and $b$ take values $0$ or  $1$, the result is independent of $a$. This implies that ${\mathbb R}_0 {\mathbb L}_0 = {\mathbb R}_0 {\mathbb L}_1$. More generally,  we have  
\begin{equation}
{\mathbb R}_0 {\mathbb L}_0 = {\mathbb R}_0 {\mathbb L}_1, ~{\mathbb R}_1 {\mathbb L}_0 = {\mathbb R}_1 {\mathbb L}_1, ~ {\mathbb L}_0 {\mathbb R}_0 = {\mathbb L}_0 {\mathbb R}_1, ~{\mathbb L}_1 {\mathbb R}_0 = {\mathbb L}_1 {\mathbb R}_1
\end{equation} 
Similarly, it is easily seen that for all  allowed values of the subcscipts  $a,b,c$
\begin{equation}
{\mathbb R}_c ~{\mathbb L}_b  ~{\mathbb R}_a = {\mathbb R}_c,~~{\mathbb L}_c ~{\mathbb R}_b~ {\mathbb L}_a = ~ {\mathbb L}_c.
\end{equation}
One can write similar equations involving longer products of the type ${\mathbb {R R L L R R}}$ etc.. This is interesting for an purely algebraic  characterization of the spider-web graphs, but it is not needed for our discussion here.

We can find eigenvectors $\psi$ that are simultaneous eigenvectors of ${\mathbb K}$ and ${\mathbb F}_i$, for all $i, 1 \leq i \leq M$.

Now, we denote the eigenvalues of ${\mathbb F}_i$ by $f_{i}$. These are $\pm 1$. Consider an eigenfunction $\psi$, with
\begin{equation}
{\mathbb F}_i \psi = f_{i} \psi, {\rm ~for~}  i = 1 {\rm ~to~} M.
\end{equation}

The Hilbert space of all  functions $\psi(m,r)$  breaks into $2^M$ disjoint sectors, each corresponding to a subspace spanned by basis vectors that are eigenvectors corresponding to a  specified set  $\{f_{i}\}$. For $M \gg N$, the number of sectors can be substantially greater than the $M 2^N$, which is the total dimension of the Hilbert space. It follows that for $M \gg N$, a large number of sectors must be null.

It is easy to see that if $\psi(m,r)$ is an eigenfunction of ${\mathbb F}_i$ with eigenvalue $-1$, then $\psi(m,r)$ must be zero, whenever $m \notin [i-N+1,i]$. If more than one of the $f_i$'s are negative, the corresponding eigenfunction $\psi(m,r)$ can be nonzero only on the intersection of the corresponding intervals. This severely constrains the allowed values of $\{f_i\}$ for a non-null sector. In particular,  this  implies that any sector with $F_i = F_j= -1$,
and distance between $i$ and $j$ greater than $N-1$ must be null.

Only in the  sector with all $f_i = +1$,  the  sector decomposition does not break translational invariance, (the sector with all $f_i = -1$ is zero-dimensional for $M > N$). In this case, we can break the sector
further into $M$ one-dimensional subsectors using the eigenvalues of ${\mathbb T}$.

It is straight forward to work out the sector decomposition and the spectrum, for small values of $N$ explicitly.

\subsection{The case $N=1$}

\begin{figure}
\centering
\includegraphics[width=4.0in]{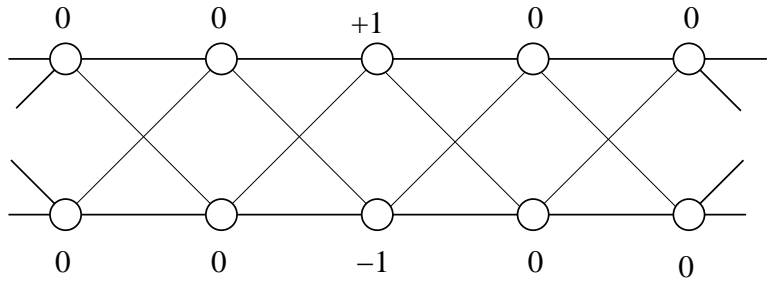}
\caption{Figure shows a part of the $N=1$ network, with   a localized eigenfunction corresponding to exactly one of the  ${\mathbb F}_i$'s negative.}
\label{fig2}
\end{figure}

For $N=1$, the graph is shown in Fig. \ref{fig2}.  There is one sector with all $f_i =+1$. This must have $\psi(m,0) = \psi(m,1)$. Thus there are $M$ linearly independent variables in this sector, and dimensionality of this sector is $M$. These are further divided into one-dimensional subsectors using the eigenvalues of ${\mathbb T}$.

Then, there are $M$ sectors with exactly one $f_i=-1$, and rest positive. In the sector with $f_1 = -1$, we must have $\psi(1,0) = -\psi(1,1)$, and $\psi(j,0) = \psi(j,1) =0 $ for $j \neq 1$. So, each of these sectors is one-dimensional.  The corresponding eigenfunction is fully localized within one level, and is shown in Fig. \ref{fig2}.  All sectors with more than one $f_i$ negative are empty. The total dimension
of the space is thus = $ 2 M$, equal to total number of modes in the problem.

\subsection{The case $N =2$}

For $N=2$, the sector with all $f_i$'s positive has dimension $M$. It is divided into $M$  one-dimensional subsectors using the eigenvalues of ${\mathbb T}$.

There are $M$ sectors where only one $f$ is negative.
Consider the sector with $f_1=-1$, and all other $f$'s positive. In this case, let the eigenfunction be $\psi(m,r)$.  We have ${\mathbb F}_1 \psi(m,r) = - \psi(m,r)$. If $m$ is not equal to $0$ or $1$, we have
${\mathbb F}_1 (m,r) =(m,r)$. This implies that $\psi(m,r)$ must be zero, for $m \neq 0,1$.  Hence the only non-zero elements are $\psi(0,0), \psi(0,1), \psi(0,2)$, $ \psi(0,3)$ and $\psi(1,0), \psi(1,1), \psi(1,2)$, $ \psi(1,3)$ . Using the symmetries of ${\mathbb F}_M$ and ${\mathbb F}_2$ we also have $\psi(0,0)=\psi(0,2)=- \psi(0,1)=- \psi(0,3)$ and $\psi(1,0)=\psi(1,1)=- \psi(1,2) =-\psi(1,3)$. Thus there are only two independent elements and hence this sector is of dimension $2$.

The number of sectors with exactly one $f_{i,\alpha}$ negative is clearly $M$. Each of these is two dimensional.

Now consider the sectors with exactly two of the $f$'s negative. Say, we have $f_i= f_j = -1$,  and the rest of the $f$ are positive. Clearly, we can assume $i \neq j$.  Using the result proved above,  if $|i-j| > 1$, then the sector is empty.

Hence, the non-trivial cases with two of the f's negative are only with $f_i =f_{i+1} = -1$, for some $i$.
Clearly, there are $M$ choices of $i$.  Say $i=1$. then the corresponding eigenfunctions can be non-zero only for sites in the level $m=1$. The values of $\psi(m =1,r)$  for different $r$ can be related to each other using the the symmetries of ${\mathbb F}_1$ and ${\mathbb F}_2$. Thus this sector is one-dimensional.

If at least three of the $f$'s are negative, then one can find a pair of values $f_i = f_j =-1$ , with $|i-j|>1$, and the sector must be empty.

Thus, we find that for $N=2$, the only non-empty sectors are those with total number of negative eigenvalues of ${\mathbb F}_i$ at most $2$.  The total number of sectors  $M + 2 M + M = 4M$, which is equal to the number of different vertices in the graph.

\subsection{Higher N}

It is straightforward to extend the treatment given above to   higher $N$. We will assume $M \gg N$, for convenience.

Given all the eigenvalues $\{f_i\}$, we can relate the value of the eigenfunction $\psi(m,r)$ for any $r$ to the value at the vertex  $(m,0)$ by applying the operators ${\mathbb F}_i$. Clearly, it differs from $\psi(m,0)$ by at most a sign, and $\psi(m,r)= \pm \phi_m$.   The dimension of the sector is the number of nonzero $\phi_m$'s that are allowed.

Consider first the  sector with all $f_i = +1$. The corresponding eigenfunction $\psi(m,r)$ must be of the form
$\psi(m,r) = g_m$, independent of $r$. There are $M$ possible choices of $m$. Hence this sector is $M$-dimensional. In this sector, ${\mathbb T}$ is a good symmetry, and we can use the the eigenvalues of ${\mathbb T}
= t_j = \exp( i \frac{2 \pi j }{M})$, with $j = 1$ to $M$, to further divide this sector into $M$ one-dimensional sectors.

Now consider the sectors with exactly one $f_i$ negative. Say $f_1 =-1$.  Then the only nonzero values of $\phi_m$ are for $m = 1, 0, -1, -2, \ldots ,2-N$. This sector is of dimension $N$.   There are $M$ such sectors.

For more than one negative  $f_i$'s, if  the largest distance between two negative $f$'s is $k$, then $k \leq N-1$. The number of non-zero $g_m$'s can be only $N-k$. Hence this is the dimension of each such sector. There are $k-1$  $f$'s between the outermost two negative $f$'s in this band, which can be $\pm 1$, and so the number of sectors with a given position of leftmost and rightmost negative $f$'s is $2^{k-1}$. also the left most negative $f$ has $M$ possible positions. Hence the number of sectors with a given value of $k$ is $M 2^{k-1}$, and the dimension of each sector is $N-k$.

As a check, this gives  the  dimension of the full Hilbert space of eigenfunctions as
\begin{equation}
M + M N + M \sum_{k=1}^{N-1}  ( N-k  ) 2^{k-1} = M 2^N
\end{equation}

\section{Spectrum of normal modes of the springs hamiltonian}

The problem of diagonalization of ${\mathbb K}$ is considerably simplified using its block-diagonalization induced by the symmetries $\{{\mathbb F}_i,{\mathbb T}\}$.  It is instructive to consider the cases of small $N$
first.

\subsection{The spectrum for  $N$=1}

In this case, ${\mathbb K}$  is fully diagonalized using the symmetries alone.  There are $2 M $ eigenstates, and the $2 M$ nontrivial sectors are all one-dimensional.  We can thus write down the eigenvectors by use of symmetries alone, and then determine their eigenvalues.

There are $M$ states in the sector with all $f_i =+1$. These are further labeled by eigenvalues of ${\mathbb T}$. The vector corresponding to the eigenvalue $e^{i k}$ of ${\mathbb T}$ is given by

\begin{equation}
\psi(m,0) = \psi(m,1) = e^{i k m},
\end{equation}

with $k = (2 \pi j )/M $, and $j$ taking integer values $0$ to $M-1$. The corresponding eigenvalue of ${\mathbb K}$ is easily seen to be

\begin{equation}
\omega^2 = 4 - 4 \cos k
\label{eq:16}
\end{equation}

Other non-null sectors have exactly one $f_i =-1$, and rest positive. The eigenvector corresponding to the $f_i =-1$ is easily written down

\begin{eqnarray}
\psi(i,0) &=& - \psi(i,1) = 1; \\ \nonumber
\psi(m,r) &=& 0, {\rm ~ for ~} m \neq i.
\end{eqnarray}

This mode is fully localized on the sites in the level $i$. The corresponding eigenvalue of ${\mathbb K}$ is seen to be $4$.

In the limit $M \rightarrow \infty$, we see that the localized modes are degenerate, and give a $\delta$- function peak in the spectrum $D(\omega^2)$ at $\omega^2 =4$. The total weight of this peak is $1/2$. The remaining modes are seen to be extended, and give rise to a continuous spectrum.  Adding the delta-function peak to the continuous part given by (\ref{eq:16}), the total density of
states is given by

\begin{equation}
D(\omega^2)= \frac{1}{2}\delta(\omega^2 -4) + \frac{1}{4 \pi [ \omega^2 (8 - \omega^2)]^{1/2}};{\rm ~ for ~}N = 1
\end{equation}

\subsection{The spectrum for $N=2$}

For the sector with  $f_i=+1$ for all $i$, again we have $M$ modes. These are given by
\begin{equation}
\psi(m,r)= \exp( i k m), {\rm ~independent ~of~} r,
\end{equation}
and the corresponding $k$ values , and the eigenvalue of ${\mathbb K}$ is as given by (\ref{eq:16}).

Now consider the sector with only one negative $f$, say $f_i = -1$, and the rest of $f$'s are all positive.
This sector is two-dimensional.
From the symmetries alone, we see that the corresponding eigenfunction $\psi(m,r)$ is non-zero only for $m=i$ and $m=i-1$. Using $f_{i-1}=f_{i+1} =+1$, we get
\begin{eqnarray}
\psi(i,00) = \psi(i,01) =-\psi(i,10) =-\psi(i,11) =g(i) \\ \nonumber
\psi(i-1,00) = - \psi(i-1,01) = \psi(i-1,10) = -\psi(i-1,11) =g(i-1)
\label{eq:20}
\end{eqnarray}
The eigenvalue equation relating $g(i-1)$ and  $g(i)$ is easily seen to simplify to
\begin{eqnarray}
- \omega^2 g(i-1) = - 4 g(i-1) + 2 g(i), \\ \nonumber
- \omega^2 g(i) = - 4 g(i) + 2 g(i-1)
\label{eq:21}
\end{eqnarray}

\begin{figure}
\centering
\includegraphics[width=4.0in]{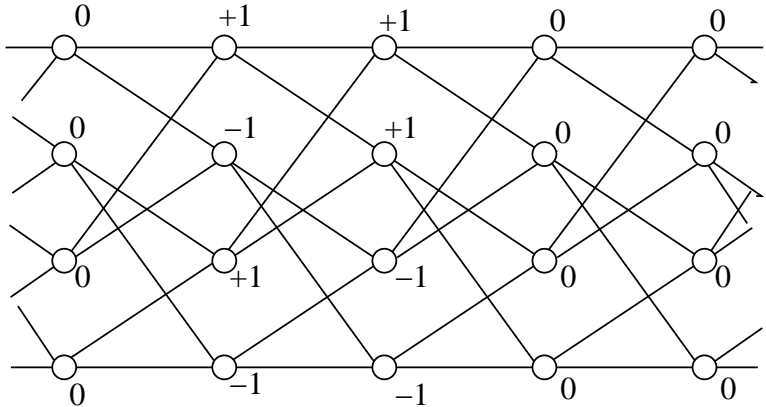}
\caption{Figure shows a part of the $N=2$ network, with   a localized eigenfunction corresponding to exactly one of the  ${\mathbb F}_i$'s negative.}
\label{fig3}
\end{figure}

This is a coupled set of two linear equations. It is easily seen that allowed solutions are $g(i) = \pm g(i-1)$, and the corresponding values of $\omega^2$ are $2$ and $6$. The eigenmode  corresponding to $\omega^2 =2$ is shown in Fig. \ref{fig3}.

For the sector with $f_i = f_{i+1}= -1$, there is only one eigenvector, which is fully localized in the level $m=i$. The corresponding eigenvalue is $4$, and the eigenvector is given by

\begin{equation}
\psi(i,00) = -\psi(i, 01) = -\psi(i,10) = \psi(i,11)
\end{equation}

In the limit $M \rightarrow \infty$, the density of states is seen to be

\begin{equation}
D(\omega^2) = \frac{1}{4}[\delta(\omega^2 -2) + \delta(\omega^2 -4) +\delta(\omega^2-6)] \\
~~~~+ \frac{1}{8 \pi [ \omega^2 (8 - \omega^2)]^{1/2}};{\rm ~ for ~}N = 2
\label{eq:22}
\end{equation}

\subsection{The spectrum for higher $N$}

The treatment given above is easily extended to higher $N$. Firstly, given the set of values $\{f_i\}$,  the corresponding eigenfunction $\psi(m,r)$ is of the form $\psi(m,r) = \pm g(m)$, and the sign is easily determined
in terms of $\{f_i\}$ and the binary representation of $r$. We choose the convention that the sign is positive for $r=0$.

\begin{figure}
\centering
\includegraphics[width=3.0in]{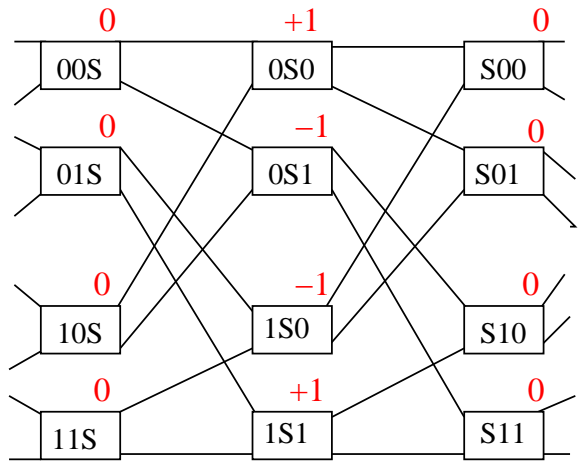}
\caption{ An eigenmode in the network with $N > 2$ with eigenvalue $4$. The amplitude, shown in red,  is nonzero  only at four sites }
\label{fig:4}
\end{figure}

A nontrivial  eigenfunction is possible, only if the indices $i$ corresponding to  negative $f$'s  occur in a narrow band of width at most $N$.  If the  width of the band is $\ell$, the number of non-zero $g(m)$ is $u =N-\ell+1$.

Consider, for simplicity, the sector in which the non-zero  $g(m)$'s
are for  $ m \in [i, i+u-1]$.  It is easily
seen that the eigenvalue equation (\ref{eq3}) in terms of the
function $g(m)$ becomes

\begin{eqnarray}
- \omega^2 g(j) &=& 2 g(j+1) + 2 g( j-1) - 4 g(j), {\rm ~for ~} j \in
[i,i +u-1], \\
\nonumber
&=& 0, {\rm ~otherwise.}
\label{eq23}
\end{eqnarray}

This equation is easily solved.  The eigenvalues are
\begin{equation}
\omega_r^2 = 4 - 4 \cos (\frac{ \pi s}{u +1}), {\rm ~for ~}s =1 {\rm ~ to ~ } u.
\label{eq:24}
\end{equation}

and the corresponding eigenfunction is
\begin{eqnarray}
g_s(m) &=& A \sin [ \frac{ \pi (m-i+1)s}{ u+1}],{\rm ~for~} m \in [ i,i +u-1],  \\
\nonumber
&=& 0,{\rm ~~otherwise}.
\label{eq25}
\end{eqnarray}

As a simple example, if $f_1 = f_N =-1$, and all $f_j= +1$, for $j > N$, then the eigenfunction  $g(m)$ is non-zero only for $m =1$, and the eigenvalue is $\omega^2 =4$. There are $2^{N-2}$ such sectors, corresponding to different possible choices of $f_i$ for $i \in [2, N-1]$.  All these different modes are localized in the level $m=1$. It is then possible to construct linear combinations of these modes, which are still eigenfunctions of ${\mathbb K}$, but no longer eigenfunctions of ${\mathbb F}_i$. We find that these can be chosen so that each eigenfunction is  highly localized, and nonzero only at four sites.

This mode is shown in Fig. \ref{fig:4}. Let $S$ be a binary string of length $N-2$. We consider the eigenfunction
which is nonzero only at the four sites $(1,0S0), (1,1S0), (1,0S1)$ and $(1,1S1)$, defined by
\begin{equation}
\psi(1,0S0)= -\psi(1,0S1) = -\psi(1,1S0) = \psi(1,1S1)
\label{eq26}
\end{equation}

Clearly, this is an eigenfunction of ${\mathbb K}$ with eigenvalue $4$.

A similar construction works for all other localized eigenmodes.  An eigenmode which extends over $u$ levels
will have  a degeneracy of $2^{N-u-1}$, and one can choose a linear combination that localizes the mode maximally. Such a mode will have non-zero amplitude only at $  2^{u+1}$ sites in each of the $u$ layers.

It is easily seen that the spectral density function for a general $N$ is:
\begin{eqnarray}
D(\omega^2)=\sum_{u=1}^{N-1} \frac{1}{2^{u+1}}\sum_{s=1}^{u} \delta(\omega^2 -(4- 4 \cos ( \frac{\pi s}{u+1}))) \nonumber \\
+\frac{1}{2^N}\sum_{s=1}^{N} \delta(\omega^2 -4 + 4 \cos [ \frac{\pi s}{N+1}]) \nonumber \\
+\frac{2^{-N}}{\pi}\sqrt{\frac{1}{\omega^2( 8-\omega^2)}}
\label{eq:28}
\end{eqnarray}

\begin{figure}
\centering
\includegraphics[width=3.0in]{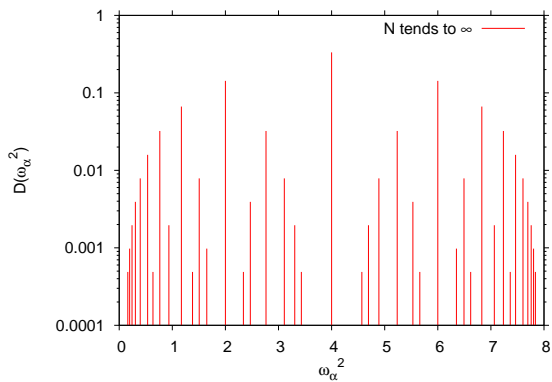}
\caption{ The spectral density in the limit of large $N$. Only the peaks with weight greater than $10^{-4}$ have been shown.}
\label{fig:5}
\end{figure}

In the limit of $M, N \rightarrow \infty$, the spectrum becomes purely discrete.  This is very interesting, as
the only other known example of a regular transitive infinite graph with a discrete spectrum of the laplacian is the Cayley graph of the lamplighter group, or its generalizations \cite{lamplighter}.

\section{Returns to the origin}

From the expression for $D(\omega^2)$, it is straightforward to determine how the probability of the random walker  being found at the starting site  at time $t$ varies with $t$ for large $t$. Using (\ref{eq:5}), and the fact that the return probability ${\rm Prob}({\bf v}_0,t;{\bf v}_0,0)$ does not depend on ${\bf v}_0$, it may be written as
\begin{equation}
{\rm Prob}({\bf v}_0,t;{\bf v}_0,0) = \int_0^{\infty}  d\omega^2  D(\omega^2) \exp(-\omega^2 t)
\end{equation}

For finite $N$, with $M$ large, the network is effectively one-dimensional. The integral above is dominated by
by the continuous spectral density for small $\omega^2$. As $D(\omega^2) \sim 2^{-N/\omega}$ for small $\omega$, it is seen that  ${\rm Prob}({\bf v}_0,t;{\bf v}_0,0)$ varies as $ 2^{-N} t^{-1/2}$ for large $t$.

For $N$ large, the spectrum is purely discrete. In (\ref{eq:24}),  For a given value of the localization width $u$, the mode with lowest frequency corresponds to $r=1$, and then for large $u$, the frequency may be
approximated by $\omega^2 \approx  2 \pi^2 u^{-2}$.  Then, we can write

\begin{equation}
{\rm Prob}({\bf v}_0,t;{\bf v}_0,0) \sim \sum_u 2^{-u} \exp( - \frac{2 \pi ^2 t}{u^2})
\end{equation}

For large $t$, the  leading behavior of this summation can be determined by steepest descent. The term corresponding to $u = u^* = \left[ \frac{ 4 \pi^2 t}{\ln 2}\right]^{1/3}$ , and we get

\begin{equation}
{\rm Prob}({\bf v}_0,t;{\bf v}_0,0) \sim  \exp \left[ - \frac{3}{2} ( 4 \pi^2)^{1/3} t^{1/3}\right],{\rm ~for~ large~}~t.
\end{equation}

We note that this is qualitatively different from the behavior for the Bethe lattice, where this probability decays exponentially with time.

The function $R(s)$ in the limit $ N \rightarrow \infty$, using (\ref{eq:7}) and (\ref{eq:28}), is given by the formula

\begin{equation}
R(s) = \sum_{u=1}^{\infty} \sum_{r=1}^{u}  \frac {1}{ 2^{u+1} ( s + 4 - 4 \cos  \frac{\pi r}{u+1})}
\end{equation}

We note that $R(s)$ can be written as
\begin{equation}
R(s) = (1/2) Q(s) -\frac{(s+4)}{s(s+8)}
\end{equation}
where
\begin{equation}
Q(s) = \sum_{u=1}^{\infty} \sum_{r=1}^{2u +1}  \frac {1}{ 2^{u+1} ( s + 4 - 4 \cos  \frac{\pi r}{u+1})}
\end{equation}

The summation over $r$ in the above equation  can be done explicitly  using  the identity
\begin{eqnarray}
\sum_{r=0}^{2u +1} \frac{1}{1 - x \cos \frac{\pi r}{u+1}} &=& \frac{2(u+1)}{ \sqrt{1- x^2}}  \left[ \frac{ (\alpha_+^{u+1} + \alpha_-^{u+1})}{(\alpha_+^{u+1} - \alpha_-^{u+1})} \right] \\
&=& \frac{T_{n+1}(1/x)}{U_n(1/x)}
\end{eqnarray}

where $\alpha_\pm = (1 \pm \sqrt{1 -x^2})/x$ and $T_{n}$ and $U_n$ are Tchebychef polynomials of the first and second kind respectively \cite{tchebychef}.

\section{Critical behavior of the spherical model}

The calculation of equilibrium properties of the spherical model is quite standard \cite{joyce}. 
We take care of the spherical constraint by using a Lagrage multiplier, and consider the Hamiltonian
\begin{equation}
{\cal{H}}_{sph} (\lambda) = {\cal{H}}_{sph} +  \frac{1}{2} \lambda \sum_{\bf v} s_{\bf v}^2 
\end{equation}
where  $\lambda$ is temperature-dependent constant, whose value is determined by the condition
\begin{equation}
 \sum_{\bf v}  \langle s_{\bf v}^2 \rangle_{\lambda}  = N_s.
\end{equation}
Here $\langle ~~\rangle_{\lambda}$ denotes equilibrium thermodynamical average using the Boltzmann factor corresponding to the Hamiltonian ${\cal{H}}_{sph} (\lambda)$.  Since ${\cal{H}}_{sph} (\lambda)$ is a purely quadratic hamiltonian, in terms of normal mode oordinates, it becomes uncoupled oscillators.  Thus, we get
\begin{equation}
\langle s_{\bf v}^2 \rangle_{\lambda} = \frac{k_B T}{N_s} \sum_{\alpha} \frac{ 1}{\lambda + \omega_{\alpha}^2} =1.
\end{equation}
 
It is easy to see that $\lambda$ is a monotonic non-decreaseing function of $T$. As the temperature is decreased from very high values, at a particular temperature $T_c$, in the thermodynamic limit $\lambda$ becomes zero, and sticks to the value zero for all $T < T_c$. The transition temperature $T_c$ is given by the equation

\begin{equation}
 k_B T_c  \int_{0}^{\infty} d\omega^2 D(\omega^2) /\omega^2 =  1
\end{equation}

For $T < T_c$, the spherical constraint  Eq. (\ref{sph_constr}), is satisfied by by a macroscopic occupation of the uniform mode. If the uniform magnetization is $m(T)$, it is easy to see that we have

\begin{equation}
m^2(T) = 1- T/T_c, {\rm ~for ~ all ~}T < T_c.
\end{equation}
 
For $T$ near $T_c$, this implies that $m(T) \sim ( T_c -T)^{1/2}$, and the magnetization exponent  $\beta$ takes the classical value $1/2$.  The internal energy per mode is the classical value $\frac{1}{2} k_B T$, by the equipartition theorem, and the specific heat takes a  constant value $(1/2)~ k_B$ for all $T < T_c$. 

For $T/T_c = 1 + \epsilon$, with $\epsilon $ small and positive, it is easy to see that $ \lambda$ is a linear function of $\epsilon$:

\begin{equation}
\lambda =  C \epsilon + {\cal{O}}({\epsilon}^{2})
\end{equation}
where C is  a positive constant. 
The specific heat for $\epsilon > 0$  is smaller than $ k_B /2$, and jumps discontinuously at $T_c$, with the discontinuity proportional to the  $C$.
 
The free energy per site $f(T)$  for $ T > T_c$ is given by the equation
\begin{equation}
f(T) = - \frac{ k_B T }{ 2 N_s} \int_{0}^{\infty} d \omega^2 D(\omega^2)  \log ( \frac{ k_B T}{ \omega^2 + \lambda})
\end{equation}

Using the fact that $ D(\omega^2)$  varies as $ exp( -C'/\omega)$ for $\omega $ tending to zero, we see that 
for small  positive $\epsilon $, $f(T)$ can be written as sum of a Taylor expandable regular part, and a singular part $ f_{sing}(T)  \sim \exp( - C'/\sqrt{\epsilon})$.

It would be interesting to extend this analysis to other  models of  phase transitions on this lattice. It seems reasonable to expect somewhat similar behavior. For example, in the  $n$-vector models with nearest -neighbor ferromagnet couplings, the high-temperature expansion for free energy involves sum over loop diagrams.
The number of self-avoiding polygons of a given perimeter $\ell$ may be expected to have a behavior similar to number of random walks that return to origin.   Note that these essential singularities can not be captured  easily in a perturbative approach.  Another interesting question is the effect of disorder on the localization properties of the normal modes of the springs Hamiltonian on this network.  These seem to be interesting directions for further study.

\ack
ACB thanks Kishore Vaigyanik Protsahan Yojana  for financial support, and TIFR for  hospitality   during his visit. DD would like to acknowledge the financial support from  the Department of Science and Technology, Government of India, through  a JC Bose Fellowship.

\end{document}